\documentclass{article}
\usepackage[margin=1in]{geometry}
\usepackage{graphicx}
\graphicspath{{BSSS_MarceloPaper_arXiv_Figs/}}
\usepackage{amsmath}
\usepackage{amssymb}
\usepackage{mathtools}
\usepackage{parskip}
\usepackage[hidelinks]{hyperref}
\usepackage{natbib}
\usepackage{subcaption}
\usepackage{color}
\usepackage{ulem}
\usepackage{booktabs}
\usepackage{amsthm}

\emergencystretch=\maxdimen
\begin{document}
\title{On minimizing cyclists' ascent times: Part~II}
\author{
	Len Bos%
	\footnote{
		Universit\`a di Verona, Italy, \texttt{leonardpeter.bos@univr.it}
	}\,,
	Michael A. Slawinski%
	\footnote{
		Memorial University of Newfoundland, Canada, \texttt{mslawins@mac.com}
	}\,,\
	Rapha\"el A. Slawinski%
	\footnote{
		Mount Royal University, Canada, \texttt{rslawinski@mtroyal.ca}
	}\,,\
	Theodore Stanoev%
	\footnote{
		Memorial University of Newfoundland, Canada, \texttt{theodore.stanoev@gmail.com}
	}
}
\date{}
\maketitle
\begin{abstract}
We formulate an optimization of a bicycle ascent time under the constraints of the average, maximum, and minimum powers.
In contrast to the first part of this study, we do not restrict the departure to flying starts with an initial speed determined by the model and its optimization.
We allow for various initial speeds, from a standstill to a launched start.
We accomplish this by generalizing the discontinuous piecewise constant speed model to a continuous piecewise linear speed model.
Regardless of the initial speed, steepness or profile of the ascent the optimal strategy tends to a constant ground speed, in agreement with the conclusion of the previous, more restricted, formulation.
This new formulation allows us to compare various initial-speed strategies and, hence, has a direct application to competitive cycling.
Notably, in timetrials composed of flat and steep sections, it helps one decide whether or not to change the bicycle, which requires stopping and restarting, from one that is more appropriate for flats to one that is more appropriate for uphills.
\end{abstract}
\section{Introduction}
In this article, we elaborate on a strategy proposed by~\citet{BosEtAl2024_STAR,BosEtAl2024_arXiv,BosEtAl2024_DRNA}, which could be viewed as Part~I of the present work, to minimize the ascent time on a bicycle, which is the goal of timetrials that include hill climbs.
In a manner similar to~\citeauthor{BosEtAl2024_DRNA}, we assume that the ascent, including distance and steepness, is known, and so is the weight of a bicycle-cyclist system, as well as fitness level, quantified by the power output, which has been the key metric in training and race preparation for the past quarter century.
Other known information consists of the air, rolling and drivetrain resistances.

In contrast to~\citet{BosEtAl2024_STAR,BosEtAl2024_arXiv,BosEtAl2024_DRNA}, we do not restrict our study to flying starts with an initial speed determined by the model and its optimization, but allow different initial speeds, including a standstill departure and a launched start, which are the two cases exemplified in this paper. 
We accomplish this by generalizing the discontinuous piecewise constant speed model of the previous paper to a continuous piecewise linear speed model.
This modification, which renders the model more flexible and empirically adequate, entails challenges in its formulation and subsequent computations.

Several other authors used similar mathematical models to conduct studies on uphill timetrialing.
\citet{AtkinsonEtAl2007} analyze variable versus constant power strategies with consideration of the cyclist power output, gradient of an ascent, and wind velocity for three race courses of varying lengths. 
\citet{Boswell2012} considers a differential equation model under the constraint of a constant-average power and considers various pacing strategies to find that substantial time savings can be realized by cyclists increasing their power on uphill sections and suitably reducing their work rates elsewhere.
Also, \citeauthor{Boswell2012} includes a derivation of an optimal pacing strategy, subject to a constant average power, to be a constant speed, which is similar to the findings of~\citet{BosEtAl2024_STAR,BosEtAl2024_arXiv,BosEtAl2024_DRNA}, but does not consider a maximum power.
\citet{RoaMunoz2018} develop a methodology to analyze the bicycle change strategy for hilly timetrial races, which involves solving an optimization problem to minimize the race time of a predetermined route as a function of the bicycle-change location along the route.
\citet{FengEtAl2022} incorporate empirical fatigue power profiles of professional cyclists and, based on the principles of mechanics and biological fatigue, solve optimization problems for the shortest competition finishing time implemented on the men 2021 UCI and women 2021 Olympic individual timetrial courses.

We begin this article by invoking a phenomenological model to estimate the power required for a given speed.
This model and its notation are similar to the ones used and discussed in several previous studies by \cite{BosEtAl2023,BosEtAl2024_SE,BosEtAl2024_STAR,BosEtAl2024_arXiv,BosEtAl2024_DRNA}.
Subsequently, we examine a numerical solution of the ascent-time minimization, and exemplify this solution for both standstill departures and launched starts.
We conclude the paper by commenting on advancements with respect to the formulation of \cite{BosEtAl2024_STAR,BosEtAl2024_arXiv,BosEtAl2024_DRNA} and on the application of this work to competitive cycling.
In the appendix, we exemplify such an application.
\section{Phenomenological power model}
\subsection{Physical considerations}
We consider a phenomenological model for a bicycle-cyclist system of mass~$m$ moving with ground speed $V$.%
\footnote{To be consistent with our previous study, \citep{BosEtAl2023,BosEtAl2024_SE}, where we distinguish between the wheel speed,~$v$\,, and the centre-of-mass speed,~$V$\,, herein we use implicitly the latter.
In this model, formally, $v\equiv V$\,; however, conceptually, we consider the motion of the centre of mass.}
The effort expended by the cyclist is quantified by power~$P=P_K+P_F$\,; the former is required to change the kinetic energy,~$\tfrac{1}{2}m\,V^2$, and the latter to overcome opposing forces, with both affected by the drivetrain resistance.
Hence,
\begin{equation}
	P
	=\dfrac{
     \overbrace{m\dfrac{{\rm d}V}{{\rm d}t}\,V}^{P_K}
    +
    \overbrace{F\,V}^{P_F}}
    {1-\lambda}\,,
\end{equation}
where $\lambda$ is the drivetrain-resistance coefficient.
$P_F$ is affected by change in elevation, as well as the resistance of rolling and of air.
Thus, explicitly,
\begin{align}
\label{eq:PV}
	P
	&=
	\quad\dfrac{
		\!\!\!\overbrace{\vphantom{\left(V\right)^2}\quad m\,\dfrac{{\rm d}V}{{\rm d}t}\quad}^\text{change in speed}
		+
        \overbrace{\vphantom{\left(V\right)^2}\,m\,g\sin\theta\,}^\text{change in elevation}
		+\
		\overbrace{\vphantom{\left(V\right)^2}
			{\rm C_{rr}}\!\!\!\underbrace{\,m\,g\cos\theta}_\text{normal force}
		}^\text{rolling resistance}
		+
		\overbrace{\vphantom{\left(V\right)^2}
			\,\tfrac{1}{2}\,\rho\,{\rm C_{d}A}\,V^{2}\,
		}^\text{air resistance}
	}{
		\underbrace{\quad1-\lambda\quad}_\text{drivetrain efficiency}
	}\,V\,,
\end{align}
where $g$ is the acceleration due to gravity, $\theta$ is the slope angle and $\rho$ is the air density.
Thus, the three model parameters are the aforementioned $\lambda$, together with the rolling-resistance coefficient,~$\rm C_{rr}$, and the air-resistance coefficient,~$\rm C_dA$.
\subsection{Power-model discretization}
In previous studies, \citet{BosEtAl2024_STAR,BosEtAl2024_arXiv,BosEtAl2024_DRNA} sought to determine the optimal strategy for minimizing the ascent time subject to the constraint of a given average power.
The hill was assumed to have no flat or descent sections, and was modelled by $N$ straight-line segments.
Each segment was traversed at a constant speed with discontinuous speed between segments.
Herein, the hill exhibits the same characteristics but the speed is represented by a piecewise linear interpolation of values at the endpoints of each segment.

\citeauthor{BosEtAl2024_DRNA} neglected the first term in the numerator of expression~(\ref{eq:PV}), which is the power,~$P_K$, associated with changes in kinetic energy.
This was justified by the fact that, riding uphill, speeds are typically low, and changes in kinetic energy are much smaller than the work done against opposing forces, especially against gravity.
Furthermore, \citeauthor{BosEtAl2024_DRNA} restricted their study to flying starts with the speed determined by the minimization of ascent time.
Under these conditions, they proved that, given an average power, the ascent time is minimized if a cyclist maintains a constant ground speed regardless of the slope.
Herein, we allow for a variable start speed, regardless of the speed that minimizes the subsequent ascent time.
Consequently, we include the first term in the numerator of expression~(\ref{eq:PV}).
As it turns out, following various start speeds, an optimal strategy still tends to a constant speed.

In accordance with expression~\eqref{eq:PV}, the power model for the cyclist can be written as
\begin{equation}
	P = \frac{1}{1-\lambda}\left(m\,\frac{{\rm d} V}{{\rm d}t}\,V+F\,V\right).
\end{equation}
Therefore, the total work done is 
\begin{equation}
\label{eq:work}
	W = \int\limits_0^TP\,{\rm d}t = \frac{1}{1-\lambda}\left(\left.\frac{1}{2}m\,V^2\right|_0^T + \int\limits_0^LF\,{\rm d}s\right),
\end{equation}
where the first term in parentheses represents the change in kinetic energy.
Note that the remaining integral contains ${\rm d}s=V{\rm d}t$.

Assuming the cyclist starts from rest, $V_1=0$,%
\footnote{This assumption is used for discussions in Sections~\ref{sub:Average} and \ref{sub:BoundedPower}, but not in Section~\ref{sub:Launched}; therein, the first term in parentheses in expression~\eqref{eq:V1} is $\tfrac{1}{2}m\left(V_T^2-V_1^2\right)$, which consequently modifies the subsequent discretization.}
we obtain
\begin{equation}
\label{eq:V1}
	W = \frac{1}{1-\lambda}\left(\frac{1}{2}m\,V_T^2 + \int\limits_0^LF\,{\rm d}s\right).
\end{equation}
Computationally, discretizing the case of $V_1=0$ is the most challenging due to the resulting discontinuity; hence, below we present it explicitly.
Computational results for other initial speeds can be obtained in a similar manner without such challenges, as we demonstrate in Section~\ref{sec:MinTime}.

Within the integrand of expression~\eqref{eq:V1}, the increment of the work done against forces opposing the motion is
\begin{equation}
	F\,{\rm d}s = \left(m\,g\,{\rm C_{rr}}\cos\theta+m\,g\sin\theta+\frac{1}{2}\,\rho\,{\rm C_dA}\,V^2\right){\rm d}s.
\end{equation}
Since  $\cos\theta\,{\rm d}s={\rm d}x$, the horizontal path increment, and $\sin\theta\,{\rm d}s = {\rm d}y$, the vertical path increment, the contributions to work from opposing gravity and rolling friction are integrable,
\begin{align}
	\nonumber\int\limits_0^LF\,{\rm d}s
	&=
	\int\limits_0^{\Delta X}m\,g\,{\rm C_{rr}}\,{\rm d}x
	+ \int\limits_0^{\Delta Y}m\,g\,{\rm d}y
	+ \int\limits_0^L\frac{1}{2}\rho\,{\rm C_dA}\,V^2\,{\rm d}s
	\\
	&=
	m\,g\,{\rm C_{rr}}\,\Delta X + m\,g\,\Delta Y + \int\limits_0^L\frac{1}{2}\rho\,{\rm C_dA}\,V^2\,{\rm d}s.
\end{align}
The only nonintegrable contribution to work comes from opposing air resistance.

To calculate this contribution, we discretize the path into $N$ straight-line segments with lengths $\{L_1,\dots,L_N\}$ and corresponding speeds $\{V_1,\dots,V_{N+1}\}$, where $V_j$ and $V_{j+1}$ are at the beginning and the end of the $j$th segment, respectively.
Then, the work done against air resistance may be written as
\begin{equation}
	\int\limits_0^L\frac{1}{2}\rho\,{\rm C_dA}\,V^2\,{\rm d}s
	=
	\frac{1}{2}\rho\,{\rm C_dA}\sum_{j=1}^NL_jV_{(j)}^2\,,
\end{equation}
where $V_{(j)}^2$ denotes the discretization of $V^2$ over the $j$th segment. 
$V_{(j)}^2\neq V_j^2$, the square of the speed at the start of the $j$th segment.
Instead, a simple discretization is 
\begin{equation}
	V_{(j)}^2 := \frac{V_j^2+V_{j+1}^2}{2},
\end{equation}
assuming $V(t)$ is continuous and piecewise linear in $t$.

Therefore, in accordance with expression~(\ref{eq:work}), the total work done\,---\,in its discretized form\,---\,is
\begin{equation}
	\label{eq:work_discr}
	W = \frac{\alpha}{2}V_{N+1}^2 + \alpha g\left({\rm C_{rr}}\Delta X+\Delta Y\right) + \beta\sum_{j=1}^NL_jV_{(j)}^2,
\end{equation}
where
\begin{equation}
	\alpha := \frac{m}{1-\lambda}
	\quad\text{and}\quad
	\beta := \frac{\tfrac{1}{2}\rho\,{\rm C_dA}}{1-\lambda}.
\end{equation}

In a similar manner, the total ascent time can be discretized as
\begin{equation}
	\label{eq:time_discr}
	T = \int\limits_0^T{\rm d}t = \int\limits_0^L\frac{{\rm d}s}{V} = \sum_{j=1}^N\frac{L_j}{V_{(j)}},
\end{equation}
where $V_{(j)}$ denotes the discretization of $V$ over the $j$th segment, with $V_{(j)}\neq V_j$, the speed at the start of the $j$th segment; instead,
\begin{equation}
	V_{(j)} := \frac{V_j+V_{j+1}}{2},
\end{equation}
assuming $V(t)$ is continuous and piecewise linear in $t$.
Along with these considerations, using work~\eqref{eq:work_discr} and time~\eqref{eq:time_discr}, the average power along the ascent is 
\begin{equation}
	\label{eq:Pavg}
	\overline{P} = \frac{W}{T}.
\end{equation}

Following the assumption that $V(t)$ is continuous and piecewise linear in $t$, the speed along the ascent is
\begin{equation}
	\label{eq:Vt_pwl}
	V(t) := V_j+M_j\left(t-t_j\right),\quad
	t_j\leqslant t <t_{j+1},
	\quad\text{for}\quad
	j = 1, \dots, N,
\end{equation}
where $M_j := (V_{j+1}-V_j)/(t_{j+1}-t_j)$, and $t_j$ and $t_{j+1}$ are the ascent times at the endpoints of the $j$th segment.

Returning to model~\eqref{eq:PV}, its discretized form can be written as
\begin{equation}
	P(t) = \left(\alpha\left(M_j+\gamma_j\right)+\beta\,V(t)^2\right)V(t),\quad
	t_j\leqslant t <t_{j+1},
	\quad\text{for}\quad
	j = 1, \dots, N,
\end{equation}
where $\gamma_j := g\left({\rm C_{rr}}\cos\theta_j+\sin\theta_j\right)$.
Since the temporal derivative of $V(t)$ and the $j$th slope angle, $\theta_j$, are discontinuous on subsequent segments, the power along the ascent is also discontinuous.
Thus, the power at the endpoints of the $j$th segment is
\begin{equation}
	\label{eq:PjPjp1}
	P(t_j) = \left(\alpha\left(M_j+\gamma_j\right)+\beta\,V_j^2\right)V_j
	\quad\text{and}\quad
	P(t_{j+1}) = \left(\alpha\left(M_j+\gamma_j\right)+\beta\,V_{j+1}^2\right)V_{j+1}.
\end{equation}
\section{Minimum ascent time}
\label{sec:MinTime}
\subsection{Preliminary comments}
According to~\citet{BosEtAl2024_STAR,BosEtAl2024_arXiv,BosEtAl2024_DRNA}, if the only constraint is a given average power, the optimal strategy for minimizing the ascent time of a cyclist is by a constant speed. 
To obtain this result, the authors neglected changes in kinetic energy and assumed that each discretized segment of the ascent is traversed at a constant speed with discontinuous speed between segments, since in the limit\,---\,as the number of segments tends to infinity\,---\,any speed can be approximated with arbitrary accuracy.
The authors also considered the case of an average-power constraint subject to a maximum power,~$P_{\rm max}$, which is a scenario that might arise if the constant-speed solution requires power along portions of the ascent that exceeds a cyclist's physical capacity. 
Hereafter, we refer to the constant-speed case as the ascent with a flying start.
In that case, the optimal strategy is to fix the speed along those portions to correspond to $P_{\rm max}$ and calculate a different, albeit constant, speed with which the rest of the ascent is completed.

In this section, we revisit the minimum-ascent-time problem, but include the power associated with changes in kinetic energy.
We assume the speed to be piecewise linear and continuous between segments.
Also, we consider the case of an average power constraint subject to power along each segment bounded by a minimum,~$P_{\rm min}$, and maximum,~$P_{\rm max}$, power.

For the numerical considerations, we use three constant-grade ascents of $\mathcal{G} = \{10\%, 20\%, 30\%\}$ such that each ascent has an elevation gain of $\Delta Y = 100\,{\rm m}$.
To achieve these grade percentages, the horizontal distance of each ascent is $\Delta X = \{1000\,{\rm m}, 500\,{\rm m}, 333.3\,{\rm m}\}$.
Following~\citet[Section~4.1]{BosEtAl2024_DRNA}, the length of each ascent is $L =\sqrt{1+\mathcal{G}^2}\,\Delta X = \{1004.9876\,{\rm m},509.9020\,{\rm m},348.0102\,{\rm m}\}$.
Also, for ease of comparison with~\cite{BosEtAl2024_STAR,BosEtAl2024_arXiv,BosEtAl2024_DRNA}, we let $m=70\,{\rm kg}$, $g=9.81\,{\rm m/s}^2$, $\rho=1.2\,{\rm kg/m}^3$, ${\rm C_dA}=0.3\,{\rm m}^2$, ${\rm C_{rr}}=0.005$, $\lambda=0.02$.
\subsection{Average-power constraint}
\label{sub:Average}
\begin{figure}
	\centering
	\includegraphics[width=0.65\textwidth]{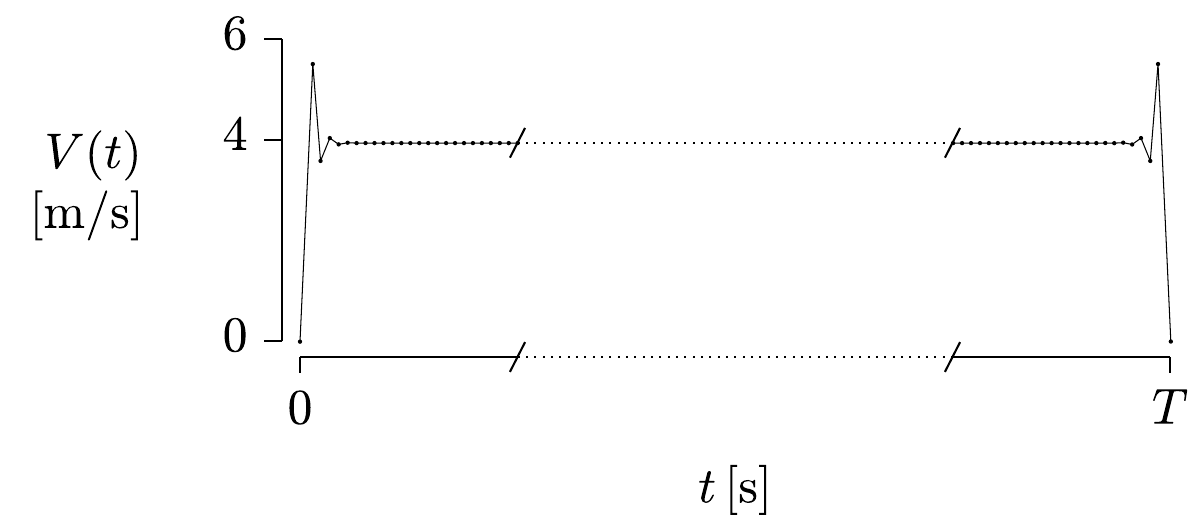}
	\caption{
		Segment speeds that minimize the ascent time along the $10$\%-grade ascent.\\
		The Heaviside step exhibits the Gibbs phenomenon due to a discretized formulation.
		The speed in the omitted region, marked by slashes on either side of the dotted line, is constant at $V=3.9441\,\rm m/s$ and corresponds to $P=300.0186\,\rm W$.
		Black dots correspond to $V_j$ and black lines to their interpolation.
	}
	\label{fig:Ascent1_VT}
\end{figure}

Using~\citet{MATLAB}'s \texttt{fmincon} optimization function, we seek to minimize the ascent time under the constraint of an average power, which we set to be~$\overline P=300\,{\rm W}$.
As shown in Figure~\ref{fig:Ascent1_VT}, the solution tends to a Heaviside step at the very beginning of the ascent, to increase the speed from zero to a constant speed of $3.9441\,{\rm m/s}$, which is maintained for the remainder of the ascent, except for another Heaviside step at the end of the ascent to return to zero.
The resulting ascent time is $T=254.8223\,{\rm s}$.
As shown in Figure~\ref{fig:Ascent1_VT}, due to sampling, the Heaviside step entails the Gibbs phenomenon.

Examining Figure~\ref{fig:Ascent1_VT}, we see that the optimal strategy is to accelerate as fast as the maximum-power constraint permits and thereafter maintain a constant speed, essentially to the end. 
The higher the maximum power, the greater the initial acceleration. 
This discontinuous behaviour creates a conflict between the true ideal optimal speed and what is permitted by the finite-dimensional piecewise linear continuous model. 
In such a case, in a manner similar to the standard Gibbs phenomenon, the optimal numerical solution exhibits  compensatory ``ringing'',  which we address in the next section.

Both the constant speed and the ascent time are consistent with the results found by \citet{BosEtAl2024_STAR,BosEtAl2024_arXiv,BosEtAl2024_DRNA}\,---\,achieved with the same discretization and average-power constraint, but with a flying start\,---\,who found the constant speed of $3.9439\,{\rm m/s}$ and the ascent time of $254.8206\,\rm s$.
However, an instantaneous acceleration is physically unreasonable; it implies an infinite power.

For $N=10\,000$ and the $10\%$-grade ascent, the Gibbs phenomenon results\,---\,at the end of the first and beginning of the last segment\,---\,in power approaching $60\,000\,{\rm W}$ and $-60\,000\,{\rm W}$, respectively.
We summarize these results, along with those for the $20\%$- and $30\%$-grade ascents, in Table~\ref{tab:avgPwr}, which, given the fine discretization of $N=10\,000$, are nearly identical to those of the flying start, found by \citet{BosEtAl2024_STAR,BosEtAl2024_arXiv,BosEtAl2024_DRNA}.
\begin{table}
	\centering
	\begin{tabular}{c*{9}{c}}
	&&&&&\multicolumn{2}{c}{Flying start}&&\multicolumn{2}{c}{Standing start} \\
	\cline{6-7}\cline{9-10}\\[-7pt]
	$\mathcal{G}\,[\%]$ & $\Delta X\,{\rm [m]}$ & $\Delta Y\,{\rm [m]}$ & $L\,{\rm [m]}$ && $V\,{\rm [m/s]}$ & $T\,{\rm [s]}$ && $V\,{\rm [m/s]}$ & $T\,{\rm [s]}$ \\
	\toprule
	$10$ & $1000$ & $100$ & $1004.9876$ && $3.9439$ & $254.8206$ && $3.9441$ & $254.8223$\\ 
	$20$ & $500$ & $100$ & $509.9020$ && $2.1174$ & $240.8104$ && $2.1176$ & $240.8107$ \\ 
	$30$ & $333\tfrac{1}{3}$ & $100$ & $348.0102$ && $1.4627$ & $237.9202$ && $1.4630$ & $237.9204$ \\ 
	\bottomrule
	\end{tabular}
	\caption{
		Ascent times, $T$, and minimizing constant speed, $V$, for three ascent profiles of increasing grade percentage, $\mathcal{G}$, under the constraint of constant-average power.
	}
	\label{tab:avgPwr}
\end{table}
\subsection{Average-power constraint subject to bounded power}
\label{sub:BoundedPower}
To render the solution physically reasonable, we constrain the power on each segment by its minimum and maximum value. 
To do so, we use expression~\eqref{eq:PjPjp1}, at the endpoints of the $j$th segment, so that
\begin{equation}
	\label{eq:bounds}
	P_{\rm min}\leqslant P_j\leqslant P_{\rm max}
	\quad\text{and}\quad
	P_{\rm min}\leqslant P_{j+1}\leqslant P_{\rm max},
	\quad\text{for}\quad
	j=1,\dots,N.
\end{equation}
Returning to~\citeauthor{MATLAB}'s \texttt{fmincon} optimization function, we seek to minimize the ascent time under the constraint of the average power,~$\overline P=300\,{\rm W}$, together with inequalities~\eqref{eq:bounds}, with $P_{\rm min} = 0\,{\rm W}$ and $P_{\rm max} = 450\,{\rm W}$.
Herein, $P=0$ corresponds to the cyclist not expending any effort whereas negative power, $P<0$, would correspond to the cyclist increasing their internal energy, which is physically unreasonable.

\begin{figure}
	\centering
	\begin{subfigure}[t]{0.65\textwidth}
		\centering
		\includegraphics[width=\textwidth]{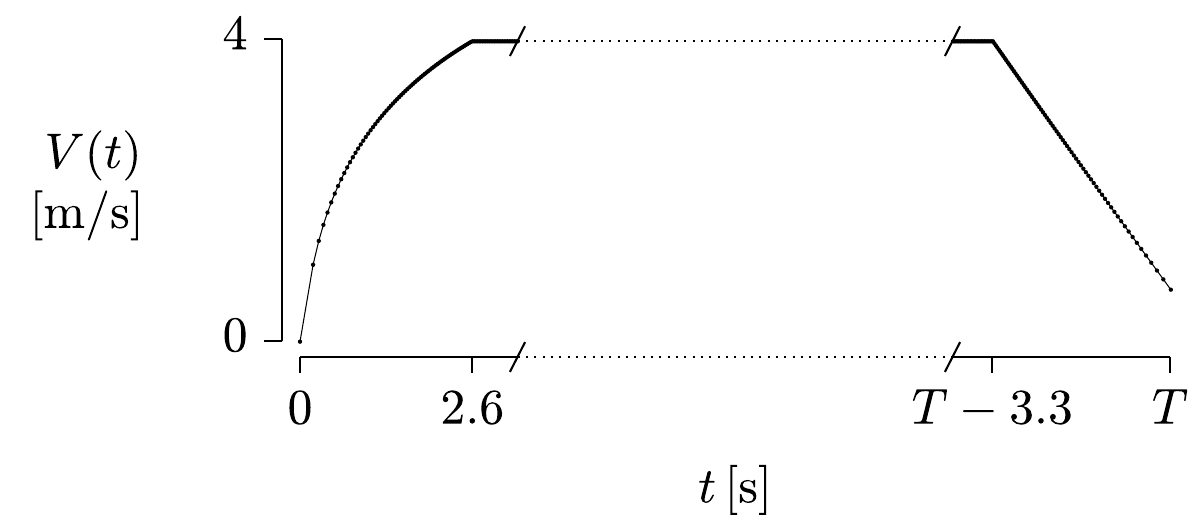}
		\caption{Speed}
		\label{subfig:VP_Ascent1_bounded_V}
	\end{subfigure}\\
	\begin{subfigure}[t]{0.65\textwidth}
		\centering
		\includegraphics[width=\textwidth]{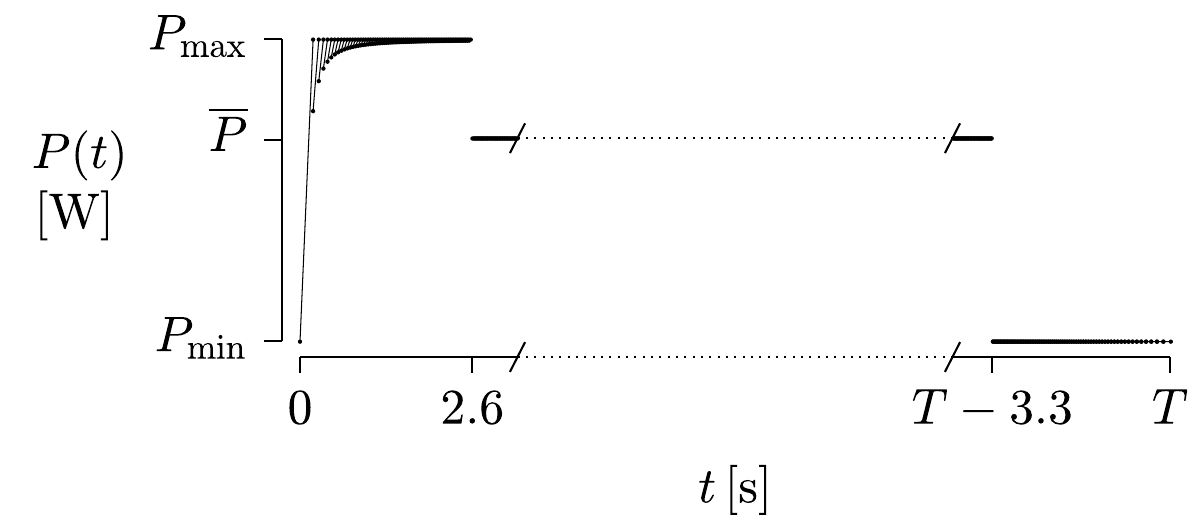}
		\caption{Power}
		\label{subfig:VP_Ascent1_bounded_P}
	\end{subfigure}
	\caption{
		Segment speed and power that yield a minimum ascent time of $T=254.9527$\,s along the 10\%-grade ascent, under the constraint of a constant-average power, $\overline{P}=300$\,W and subject to power bounded by $P_{\rm min}$ and $P_{\rm max}$.
		The speed in the omitted region, marked by slashes on either side of the dotted line, is constant at $V=3.9811\,\rm \,m/s$ and corresponds to $P=302.6858\,\rm W$.
		Black dots correspond to $V_j$ in (\subref{subfig:VP_Ascent1_bounded_V}), but $P_j$ and $P_{j+1}$ in (\subref{subfig:VP_Ascent1_bounded_P}); black lines correspond to the interpolation of each data point's respective endpoints.
	}
	\label{fig:VP_Ascent1_bounded}
\end{figure}

Choosing the $10\%$-grade ascent and setting $N=10\,000$, we find that the the minimum ascent time is $T=254.9527\,{\rm s}$, obtained with a constant speed of $V=3.9767\,{\rm m/s}$.
At the beginning of the ascent, it requires $2.5986\,{\rm s}$ to increase the speed from a standstill to the constant speed without exceeding~$P_{\rm max}$.
At the end of the ascent, it requires $3.3258\,{\rm s}$ to decrease the speed from the constant speed to a final speed of $V_{N+1}=0.5175\,{\rm m/s}$ without  reaching negative powers.
The segment speed and power are shown in Figure~\ref{fig:VP_Ascent1_bounded}.
These results, along with those for the $20\%$- and $30\%$-grade ascents, are summarized in Table~\ref{tab:avgPwr_bounded}.

The deceleration at the end of the ascent found in the numerical solution and shown in Figure~\ref{fig:Ascent1_VT} may appear surprising, since it results in a longer ascent time than maintaining a constant speed would.
However, it is intuitively understandable. 
Since the solution minimizes the ascent time $T$, the average power constraint in equation~\eqref{eq:Pavg} implies that the total work $W$ is also minimized.
The numerical solution accomplishes this objective by minimizing $\Delta K$, which, since herein the initial speed is taken to be zero, implies a final speed that approaches zero.
In physical terms, maintaining a substantial speed at the end of the ascent would result in greater than optimal work being performed.
\begin{table}
	\centering
	\begin{tabular}{c*{12}{c}}
	&&\multicolumn{2}{c}{Flying start}&&\multicolumn{2}{c}{Standing start}&&\multicolumn{2}{c}{Launched start} \\
	\cline{3-4}\cline{6-7}\cline{9-10}\\[-7pt]
	$\mathcal{G}\,[\%]$ && $V\,{\rm [m/s]}$ & $T\,{\rm [s]}$ && $V\,{\rm [m/s]}$ & $T\,{\rm [s]}$ && $V\,{\rm [m/s]}$ & $T\,{\rm [s]}$\\
	\toprule
	$10$ && $3.9441$ & $254.8223$ && $3.9811$ & $254.9527$ && $4.0977$ & $240.9870$ \\ 
	$20$ && $2.1176$ & $240.8107$ && $2.1250$ & $240.8172$ && $2.1707$ & $225.7489$ \\ 
	$30$ && $1.4630$ & $237.9204$ && $1.4652$ & $237.9213$ && $1.4888$ & $222.4865$ \\ 
	\bottomrule
	\end{tabular}
	\caption{
		Ascent times, $T$, and minimizing constant speed, $V$, for three ascent profiles of increasing grade percentage, $\mathcal{G}$, under the constraint of constant-average power and subject to power bounded by $P_{\rm min}$ and $P_{\rm max}$.
		Initial speed of the launched efforts is $V_1=11.7217\,{\rm m/s}$, which corresponds to $P=300\,{\rm W}$ on a flat section.
	}
	\label{tab:avgPwr_bounded}
\end{table}
\subsection{Launched start}
\label{sub:Launched}
Let us consider a case for which the cyclist approaches the ascent along a horizontal road.
Given the power along that road that is equal to the average-power constraint,~$\overline P$, along the hill, the speed at the beginning of the ascent is greater than the subsequent ascent speed.
Hence, we refer to such a scenario as a launched start.
Notably, such situations are common in competitions, where the last part of the timetrial is an ascent to the finish line.

\begin{figure}[h]
	\centering
	\includegraphics[width=0.65\textwidth]{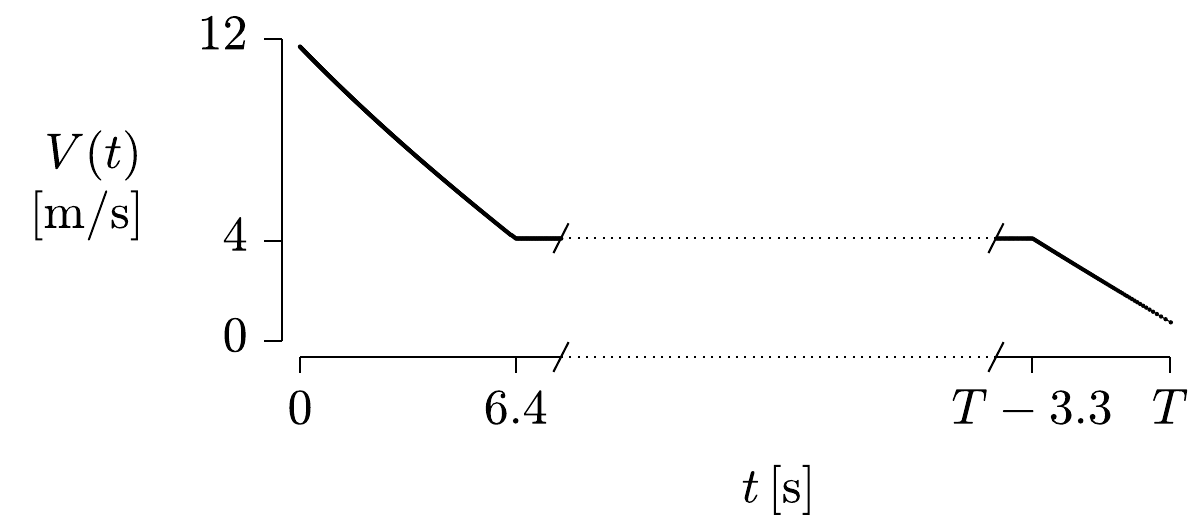}
	\caption{
		Segment speeds that yield a minimum ascent time of $T=240.9870\,{\rm s}$ in a launched effort along the 10\%-grade ascent, under the constraint of a constant-average power, $\overline{P}=300$\,W and subject to power bounded by $P_{\rm min}$ and $P_{\rm max}$.
		The speed in the omitted region, marked by slashes on either side of the dotted line, is constant at  $V=4.0977\,\rm m/s$ and corresponds to $P=312.6285\,\rm W$.
}
	\label{fig:VP_Ascent1_bnd_lnchd}
\end{figure}

Letting $P=300\,\rm W$ and $\theta=0$ in expression~\eqref{eq:PV}, we obtain, through algebraic manipulation, the depressed cubic equation,
\begin{equation}
	V^3 +\frac{\alpha\,{\rm C_{rr}}}{\beta}V-\frac{P}{\beta}=0\,;
\end{equation}
herein, the solution is $V=11.7217\,{\rm m/s}$.
Using this value as the initial uphill speed, with $\mathcal{G}=10\%$, $\overline P=300\,\rm W$, $P_{\rm min}=0\,{\rm W}$, and $P_{\rm max}=450\,{\rm W}$, the minimum ascent time is $T=240.9870\,{\rm s}$. 
As illustrated in Figure~\ref{fig:VP_Ascent1_bnd_lnchd}, the solution tends to a constant speed of $4.0977\,{\rm m/s}$, which corresponds to a power of $312.6285\,{\rm W}$, remains at that speed for the majority of the ascent, and decreases to a final speed of $V_{N+1}=0.6185\,{\rm m/s}$. 
The initial decrease in speed takes $6.4175\,{\rm s}$ whereas the final decrease takes $3.3408\,{\rm s}$.
These results, along with those for the $20\%$- and $30\%$-grade ascents, are summarized in the ``Launched start'' column of Table~\ref{tab:avgPwr_bounded}.
Since $V_1\neq0$, the first term in parentheses in expression~\eqref{eq:V1} is $\tfrac{1}{2}m\left(V_T^2-V_1^2\right)$, which consequently modifies the subsequent discretization.

One of the main applications of a launched-start model is to gain an insight into timetrials composed of flat and steep sections by examining a strategy of stopping before the ascent to change a bicycle that optimizes a flat course, where the effort is dominated by overcoming the air resistance, to a bicycle for an ascent, where the effort is dominated by overcoming the gravity.
Comparing the ascent times in Table~\ref{tab:avgPwr_bounded} for the standing and launched starts, we notice a significant time loss due to stopping and restarting.
One needs to decide whether or not such a loss is to be more than compensated by a choice of a more appropriate bicycle.

Let us consider a flat course ending with a $1004.9876\,\rm m$ ascent of~10\%.
For the flat portion of the race, the cyclist uses a bicycle that results in $m = 70\,{\rm kg}$; the bicycle has disk wheels and handlebar extensions to minimize the air resistance.
Hence, we set the resulting air-resistance coefficient to ${\rm C_dA} = 0.2\,{\rm m}^2$.
Letting $P=300\,{\rm W}$, $\theta=0$, $g=9.81\,{\rm m/s}^2$, $\rho=1.2\,{\rm kg/m}^3$, ${\rm C_{rr}}=0.005$ and $\lambda=0.02$, in equation~(\ref{eq:PV}), we obtain $V=13.4089\,{\rm m/s}$.

If there is no bicycle change, the cyclist starts the ascent with $V_1=13.4089\,{\rm m/s}$.
Since the position for an ascent is less aerodynamic, we set ${\rm C_dA} = 0.3\,{\rm m}^2$.
The resulting ascent time is $T = 237.3167\,{\rm s}$.

If the cyclist changes for a lighter bicycle that results in $m = 68\,{\rm kg}$, there is a need to restart from \mbox{$V_1=0\,\rm m/s$}.
Again, for a less aerodynamic climbing position, we set ${\rm C_dA} = 0.3\,{\rm m}^2$.
The resulting ascent time is \mbox{$T = 248.4836\,{\rm s}$}.

Comparing these ascent times, we see that, in this case as well as for the two other cases shown in Table~\ref{tab:bikeChange}, restarting with a lighter bicycle is not sufficiently beneficial; launched starts still result in shorter times.
Nevertheless, comparing the ascent time for standing starts in Table~\ref{tab:avgPwr_bounded} with corresponding times in Table~\ref{tab:bikeChange}, we notice a significant, albeit overall insufficient, improvement with a lighter bicycle.

The formulation presented herein allows us to adjust input parameters for a problem at hand to help in choosing a race strategy by quantifying the wisdom of an adage that the longer and steeper the ascent the more likely one benefits from a bicycle change; we exemplify such an approach in Appendix~\ref{app:CBS}.
To arrive at a final decision, a competitor also needs to take into account time spent for such actions as slowing down at the end of a flat section and the bicycle change itself, which are not included in this formulation. 
\begin{table}
	\centering
	\begin{tabular}{c*{12}{c}}
	&&\multicolumn{2}{c}{Standing start}&&\multicolumn{2}{c}{Launched start} \\
	&&\multicolumn{2}{c}{($m=68\,{\rm kg}$)}&&\multicolumn{2}{c}{($V_1=13.4089\,{\rm m/s}$)} \\	
	\cline{3-4}\cline{6-7}\\[-7pt]
	$\mathcal{G}\,[\%]$ && $V\,{\rm [m/s]}$ & $T\,{\rm [s]}$ && $V\,{\rm [m/s]}$ & $T\,{\rm [s]}$ \\
	\toprule
	$10$ && $4.0920$ & $248.4836$ && $4.1246$ & $237.3167$ \\ 
	$20$ && $2.1884$ & $234.0619$ && $2.1787$ & $221.5468$ \\ 
	$30$ && $1.5094$ & $231.1644$ && $1.4929$ & $218.0571$ \\ 
	\bottomrule
	\end{tabular}
	\caption{
		Ascent times, $T$, and minimizing constant speed, $V$, for three ascent profiles of increasing grade percentage, $\mathcal{G}$, under the constraint of constant-average power and subject to power bounded by $P_{\rm min}$ and $P_{\rm max}$.
		Mass of the standing start efforts is $m=68\,{\rm kg}$.
		Initial speed of the launched efforts is $V_1=13.4089\,{\rm m/s}$, which corresponds to $P=300\,{\rm W}$ on a flat section with ${\rm C_dA}=0.2\,{\rm m}^2$.
	}
	\label{tab:bikeChange}
\end{table}
\section{Conclusion}
Generalizing the formulation of \cite{BosEtAl2024_STAR,BosEtAl2024_arXiv,BosEtAl2024_DRNA} by not restricting the initial speed to a value determined by the model and its optimization, we gain an insight into the optimization process and its results.
Similarly to the result proven analytically by \citeauthor{BosEtAl2024_DRNA} in the aforementioned formulation, the minimization tends to a constant speed for the ascent regardless of the initial speed and the ascent profile.
Notably, as stated in Tables~\ref{tab:avgPwr} and \ref{tab:avgPwr_bounded}, the ascent times are almost identical to those in \citeauthor{BosEtAl2024_DRNA}

In the present paper, for conciseness and ease of comparison, we use only constant-grade ascents.
However, we have confirmed by numerical experimentation that the minimization tends to a constant speed for arbitrary ascent profiles, such as the ones discussed by \cite{BosEtAl2024_DRNA}, and shown in their Figures~4(b),\,4(c),\,4(d),\,4(e).
This remains true for models subject to the average-power, minimum-power and maximum-power constraints.
If the two latter constraints are included, a resulting constant speed is different from the one under the average-power constraint alone to accommodate the parts of the ascent that are restricted by the power bounds.

This generalization also renders the model more flexible and empirically adequate.
As discussed in Section~\ref{sec:MinTime}, the model is ready to be used for uphill timetrials and for timetrials composed of flat and steep sections.
Future work might focus on presenting strategies for specific races and examining their results.
\section*{Acknowledgements}
We wish to acknowledge insightful editorial comments and proofreading of Scott Anderson and David Dalton as well as artistic contributions of Roberto Lauciello.
Also, since phenomenological models require an evaluation of their empirical adequacy, we would like to acknowledge the cycling team of Matteo Bertrand and Alberto Demicheli, guided by G.\,Andrea Oliveri.
The research is partially supported by NSERC Discovery Grant RGPIN-2018-05158 of Michael A. Slawinski.
\bibliographystyle{apa}
\bibliography{BSSS_MarceloPaper_arXiv.bib}
\begin{appendix}
\setcounter{figure}{0}
\renewcommand{\thefigure}{\thesection.\arabic{figure}}
\renewcommand{\theequation}{\Alph{section}.\arabic{equation}}
\section{Change-of-bicycle strategy}
\setcounter{equation}{0}
\label{app:CBS}
Comparing the ascent from the standstill start, examined in Section~\ref{sub:BoundedPower}, with the ascent from the launched start, examined in Section~\ref{sub:Launched}, we address the change-of-bicycle strategy for timetrials composed of a flat portion followed by a steep one.

\begin{figure}[h]
\centering
\includegraphics[width=0.5\textwidth]{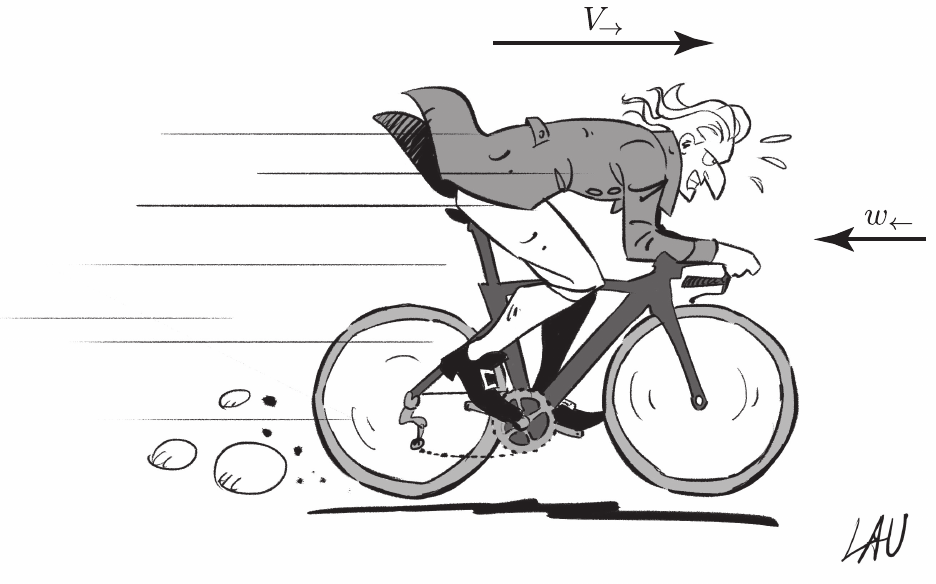}
\caption{Lagrange on an aerodynamic bicycle}
\label{fig:Lagrange}
\end{figure}

\begin{figure}[h]
\centering
\includegraphics[width=0.5\textwidth]{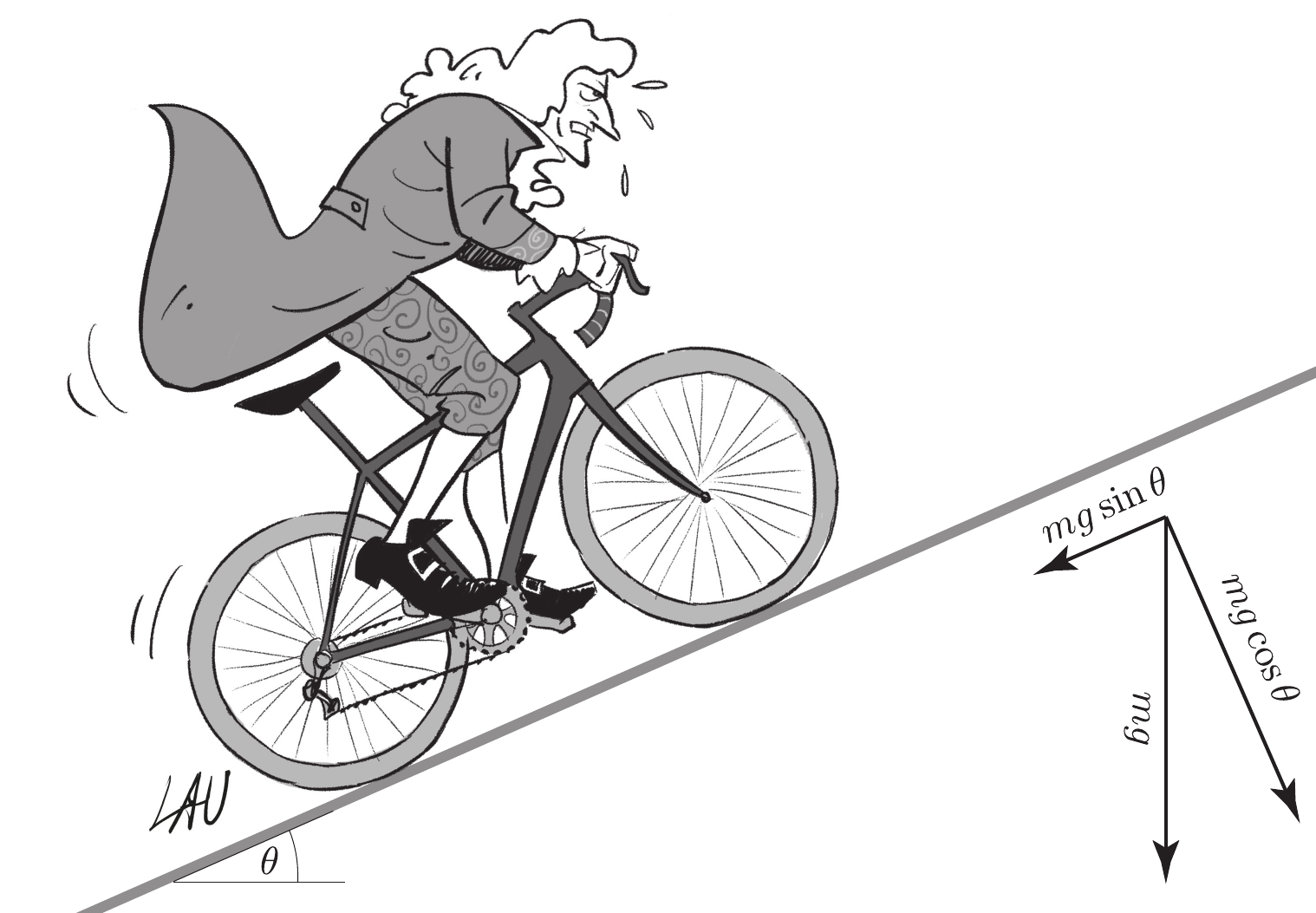}
\caption{Newton on a light uphill bicycle}
\label{fig:Newton}
\end{figure}

Along the flats, the cyclists' effort is dominated by overcoming air resistance.
Hence, a bicycle is equipped with disk wheels and handlebar extensions to minimize that resistance, as illustrated in Figure~\ref{fig:Lagrange}.
Along a steep uphill, the cyclists' effort is dominated by overcoming gravity.
Hence, the strategy is to use a bicycle that is as light as possible, as illustrated in Figure~\ref{fig:Newton}.
For instance, disk wheels of the former are heavier than spoked wheels of the latter.
Thus, we have the following strategic question: should the cyclist change the bicycle at the beginning of the climb, which requires stopping and restarting?

Let us consider a scenario described in Section~\ref{sub:Launched}.
In particular, we choose a constant%
\footnote{As shown by \citet[Table~1]{BosEtAl2024_arXiv}, the minimized ascent times for constant-grade uphills are almost equal to the minimized times of varied-grade uphills, as long as the elevation gains are the same.
Hence, strategic insights presented below are also valid for a given average grade.}
grade of $\mathcal{G}=10\%$, and we allow for a variable ascent length,~$L$\,.
We examine the minimum ascent time along the inclined section, under the constraint of average power,~$\overline{P}=300\,{\rm W}$, subject to $P_{\rm min}=0\,{\rm W}$ and $P_{\rm max}=450\,{\rm W}$. 
We assume that\,---\,along the flats\,---\,the cyclist uses an aerodynamic bicycle that results in $m=70\,{\rm kg}$ and ${\rm C_dA}=0.2\,{\rm m}^2$.
For the uphill, in a less aerodynamic position, we set ${\rm C_dA}=0.3\,{\rm m}^2$.
We consider two scenarios.
\begin{enumerate}
	\item
	The cyclist does not change bicycles and starts the ascent with an initial speed of $V_1=13.4089\,\rm m/s$, which is the speed obtained using equation~\eqref{eq:PV} with $P=300\,{\rm W}$ and $\theta=0$\,.
	\item
	The cyclist changes for a lighter bicycle that results in $m=68\,{\rm kg}$ and starts the ascent from a standstill,~$V_1=0\,{\rm m/s}$.
\end{enumerate}
We assume that the rolling resistance, ${\rm C_{rr}}=0.005$, and drivetrain resistance, $\lambda=0.02$, are the same for both bicycles.

\begin{figure}[h]
	\centering
    \includegraphics[width=0.8\textwidth]{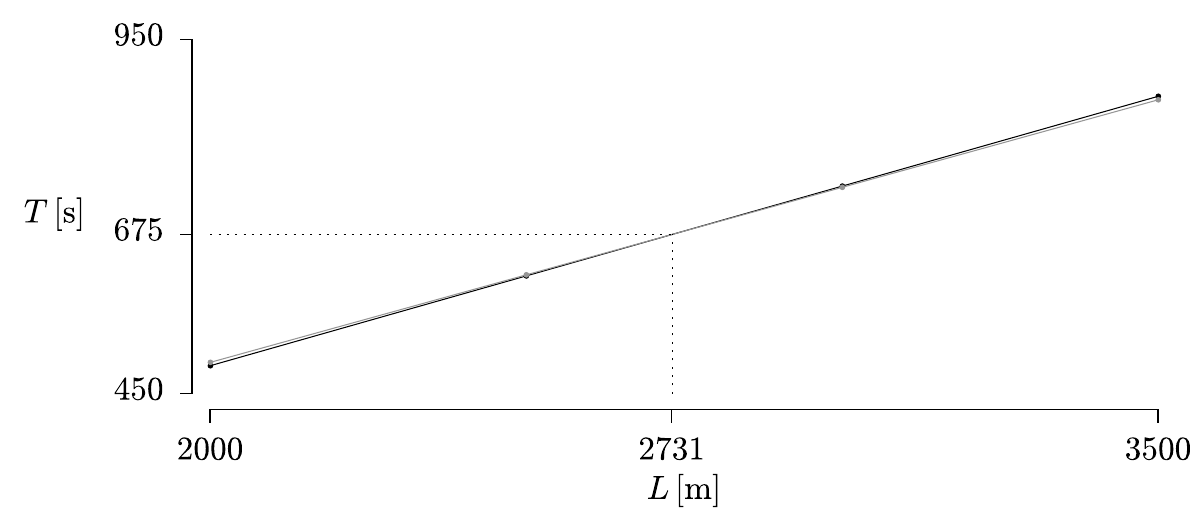}
    \caption{Minimum ascent times: for the launched bicycle in black and for the changed one in grey}
\label{fig:ChngBk_TvsL}
\end{figure}

We use \citeauthor{MATLAB}'s \texttt{fmincon} optimization function and perform the ascent-time minimization for ascent lengths ranging from $L=250\,{\rm m}$ to $L=10\,000\,{\rm m}$, with $500\,{\rm m}$ increments from $L=500\,{\rm m}$.
In Figure~\ref{fig:ChngBk_TvsL}, we present the minimum ascent times as a function of distance.

As expected, initially the ascent time is shorter for the launched bicycle.
From a certain point on, however, the ascent time is shorter for the lighter bicycle.
To find that point, we recognize a linear relation between time and distance and perform the least-squares regression.
For the first scenario,
\begin{equation}
\label{eq:AeroTL}
T=0.2536\,L-17.4650\,;
\end{equation}
for the second,
\begin{equation}
\label{eq:LightTL}
T=0.2471\,L+0.1308\,.
\end{equation}
Using these results, we obtain the intersection point,~$(L,T)=(2730.5220\,{\rm m}\,,674.8639\,{\rm s})$.

In both regressions, the coefficient of determination is $R^2=1$\,: the points are aligned perfectly, which is a supporting evidence for an extension of the proof presented by \citet[Section~3]{BosEtAl2024_arXiv}: given an average power, the ascent time is minimized if a cyclist maintains a constant ground speed regardless of the slope and\,---\,as inferred herein from regression analysis\,---\,regardless of the initial speed.
The reciprocals of slopes in equations~(\ref{eq:AeroTL}) and (\ref{eq:LightTL}) are the estimated ground speeds: $3.9432\,\rm m/s$, for the launched bicycle, and $4.0469\,\rm m/s$, for the lighter one.

\begin{figure}[h]
\centering
\includegraphics[width=0.8\textwidth]{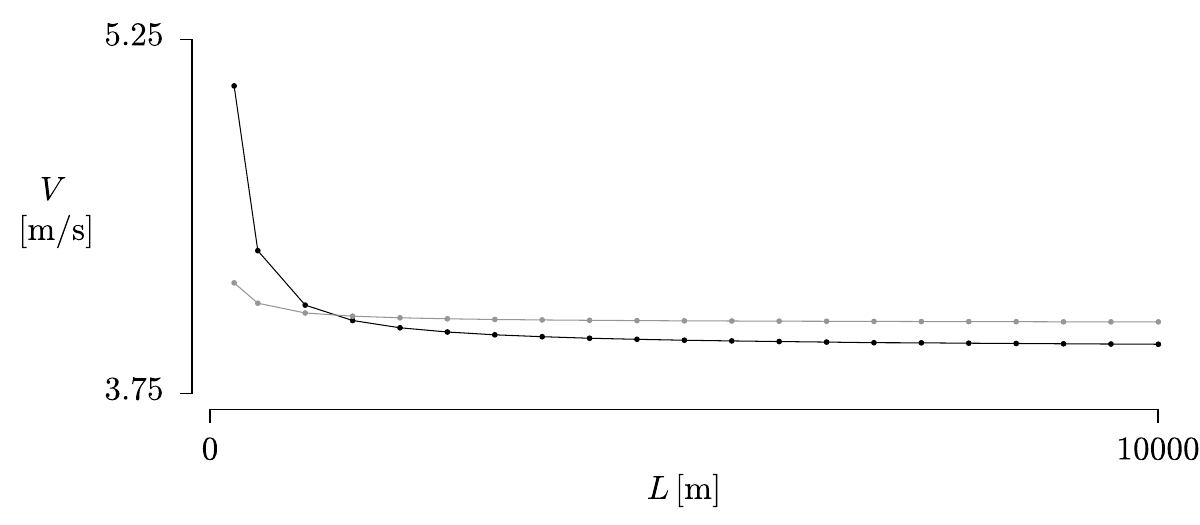}
\caption{Optimal speeds as a function of distance: for the launched bicycle in black and for the changed one in grey}
\label{fig:ChngBk_VvsL}
\end{figure}

In Figure~\ref{fig:ChngBk_VvsL}, we present speeds along the uphill.
We notice that after about a thousand meters, in spite of the initial launch, the extra $2\,{\rm kg}$ brings the speed of the launched bicycle below the speed of the lighter one.
As expected, in either scenario, $V$ approaches a constant speed.

In conclusion\,---\,for the example discussed herein, and taking into account slowing down at the end of a flat section and the time of changing the bicycle\,---\,we might consider the second scenario to be beneficial for uphills longer than about three kilometres.
In general, the steeper the uphill, the greater the effect of the weight difference; hence, even a shorter but steeper uphill justifies a bicycle change.
For instance, for a $20\%$ uphill, the regression lines analogous to equations~(\ref{eq:AeroTL}) and (\ref{eq:LightTL}) are
\begin{equation*}
T = 0.4723\,L - 19.2716
\end{equation*}
and
\begin{equation*}
T = 0.4590\,L + 0.0060\,,
\end{equation*}
respectively, whose intersection point is $(L,T)=(1454.1236\,{\rm m}\,,667.4764\,{\rm s})$, and the ground speeds are $2.1173\,\rm m/s$ and $2.1786\,\rm m/s$.

In the above discussion, we implicitly assume that the flat section is sufficiently long not to question the use of an aerodynamic bicycle.
There are, however, cases for which a flat section is short enough to use an uphill bicycle from the very start.

To gain an insight into this question, we set~$\theta=0$ in model~\eqref{eq:PV}.
Following Section~\ref{sec:MinTime}, we let $m=70\,{\rm kg}$ and ${\rm C_dA}=0.2\,{\rm m}^2$, for an aerodynamics bicycle, and $m=68\,{\rm kg}$ and ${\rm C_dA}=0.3\,{\rm m}^2$, for an uphill one.
Other values are the same as in that section, including the average power of~$\overline{P}=300\,{\rm W}$.
 Solving for speed, we obtain $V_1=12.7742\,{\rm m/s}$ and $V_2=11.2526\,{\rm m/s}$ for the aerodynamic and uphill bicycles, respectively.
 
The time gained on an aerodynamic bicycle is
\begin{equation}
\label{eq:DeltaT}
\Delta T=L\,\dfrac{V_1-V_2}{V_1\,V_2}=0.0106\,L\,,
\end{equation}
where $L$ is the length of a flat section.
Given~$L$\,---\,and if the uphill section is sufficiently steep and long to justify a lighter bicycle\,---\,examining the values of~$\Delta T$ helps us decide whether or not to start with an aerodynamic bicycle.
Also, we can solve equation~(\ref{eq:DeltaT}) for~$L$, set $\Delta T$ to the value that justifies the use of an aerodynamic bicycle, and, hence, find the required length of the corresponding flat section.
\end{appendix}
\end{document}